\PassOptionsToPackage{xcolor}{dvipsnames,svgnames,x11names}
\documentclass[twocolumn]{aastex631}

\usepackage{amsmath}
\usepackage[varg]{txfonts}
\usepackage{algorithmic}
\usepackage{multirow}
\usepackage[lined,ruled,linesnumbered]{algorithm2e}
\usepackage{graphicx}
\usepackage{tikz}
\usetikzlibrary{shapes.geometric,shapes.arrows,decorations.pathmorphing,
decorations.markings,decorations.pathreplacing}
\usetikzlibrary{matrix,chains,scopes,positioning,arrows,fit,spy,calligraphy}  

\usepackage{savesym}
\savesymbol{listofchanges}

\usepackage[authormarkup=none, commandnameprefix=always, final]{changes}

\newcommand{\stkout}[1]{\ifmmode\text{\sout{\ensuremath{#1}}}\else\sout{#1}\fi}
\setdeletedmarkup{}

\begin{document}
\definechangesauthor[color=purple]{rev}
\definechangesauthor[color=teal]{min}
\definechangesauthor[color=OliveGreen]{amin}

\title{A Decentralized Framework for Radio-Interferometric Image Reconstruction}

\correspondingauthor{S. Wang}
\email{sunrise.wang@oca.eu}

\author[0000-0002-5038-9531]{S. Wang}
\affiliation{Universit\'e C\^ote d’Azur, Observatoire de la C\^ote d’Azur, CNRS, 06000 Nice, France}

\author[0000-0002-8515-939X]{S. Mignot}
\affiliation{Universit\'e C\^ote d’Azur, Observatoire de la C\^ote d’Azur, CNRS, 06000 Nice, France}

\author[0000-0002-1755-4582]{S. Prunet}
\affiliation{Universit\'e C\^ote d’Azur, Observatoire de la C\^ote d’Azur, CNRS, 06000 Nice, France}

\author[0000-0003-3586-4485]{L. Di Mascolo}
\affiliation{Kapteyn Astronomical Institute, University of Groningen, Landleven 12, 9747 AD, Groningen, The Netherlands}
\affiliation{Universit\'e C\^ote d’Azur, Observatoire de la C\^ote d’Azur, CNRS, 06000 Nice, France}

\author[0000-0003-0148-3254]{M. Spinelli}
\affiliation{Universit\'e C\^ote d’Azur, Observatoire de la C\^ote d’Azur, CNRS, 06000 Nice, France}

\author[0000-0002-7297-7625]{A. Ferrari}
\affiliation{Universit\'e C\^ote d’Azur, Observatoire de la C\^ote d’Azur, CNRS, 06000 Nice, France}

\begin{abstract}
The advent of large aperture arrays, such as the ones currently under construction for the SKA project, allows for observing the Universe in the radio-spectrum at unprecedented resolution and sensitivity. To process the enormous amounts of data produced by these telescopes, scalable software pipelines are required. This paper helps address this by proposing a framework that allows for decentralized radio-interferometric image reconstruction, parallelizing by spatial frequency. This is achieved by creating pseudo-full-resolution problems for each node by using the local visibilities together with previous major cycle reconstructed images from the other nodes. We apply the proposed framework to both multiscale CLEAN and sparsity regularized convex reconstruction and compare them to their serial counterparts across four different data sets of varying properties in the context of two visibility partitions. We found that the parallelization framework allows for significantly improved reconstruction times for images of similar quality. This was especially the case for our larger datasets where we were able to achieve close to the optimal $2\times$ speedup. 
\end{abstract}


%--------------------------------------------------------------

\section{Introduction} 
\label{S:intro}
The two telescopes currently under construction for The SKA project\footnote[3]{\url{https://www.skao.int/}}, SKA-Mid ($350$~MHz to $15.35$~GHz) in South Africa and SKA-Low ($50-350$ MHz) in Australia, will allow us to observe the Universe \chreplaced[id=rev]{at}{within} these frequencies at unprecedented resolutions and sensitivities. These more detailed observations will result in large amounts of data to be processed. For SKA-Mid, projections estimate up to $2.375\;\text{TB/s}$ from the dishes to the beamformer and correlator engines, and $1.125\;\text{TB/s}$ from these to the imaging supercomputer~\citep{skamid-spie22}. For SKA-Low, the estimated data transfer is $0.725\;\text{TB/s}$ from the antennas to the correlator and $0.29\;\text{TB/s}$ from the correlator to the imaging supercomputer~\citep{skalow-spie22}.

With these data rates, it is imperative for current image reconstruction algorithms to scale. On a coarse \chreplaced[id=min]{level}{scale}, this involves parallelizing the sample data (i.e. visibilities) processing across multiple nodes of a cluster. This is typically performed along the embarrassingly parallel frequency and time domains, with independent reconstruction of each partition. Approaches that separate the image into facets for direction-dependent calibration, such as~\citet{cornwell1992radio,van2016lofar,tasse2018faceting}, also enable parallelization in the spatial domain. This is more complicated due to the corner turn caused by each facet depending on all visibilities and vice versa\chadded[id=rev]{, but there has been recent work by~\citet{Wortmann_Kent_Nikolic_2024} that looks to address this}.

Another possible axis to parallelize is by spatial frequency. This requires more careful treatment compared to parallelizing by time and frequency, since image reconstruction methods typically need to process all the \chadded[id=rev]{spatial} frequencies together in order to obtain full-resolution reconstructions, leading to only a modicum of work being done in this respect~\citep{onose2016scalable, pratley2019distributed, van2018image, Wortmann_Kent_Nikolic_2024}. This paper addresses this by proposing a general decentralized framework to parallelize radio-interferometric image reconstruction by spatial frequency. This is achieved by filling in the missing information in each node with reconstructed images from the others, allowing each node to simultaneously reconstruct full-resolution images. It then combines these to obtain the final reconstructed image.

We present our framework within the traditional major-minor loop image reconstruction paradigm, and apply it to two different image reconstruction methods, one based on convex optimization with sparsity regularization, such as in~\citet{wiaux2009compressed, carrillo2014purify, garsden2015lofar, dabbech2015moresane}, and the other being multi-scale CLEAN~\citep{cornwell2008multiscale}. We evaluated our framework by comparing these parallel implementations with their serial counterparts across four different datasets. We found our framework to produce reconstructed images of comparable quality in substantially less time than the serial approaches, particularly in the case of larger datasets.

The remainder of this paper is structured as follows: We provide a brief overview of radio-interferometric imaging as well as a review of the literature in Section~\ref{S:RadioInterferometry}; we describe our proposed parallelization framework in Section~\ref{S:Parallelization}; we discuss how it can be applied to two existing reconstruction methods in Section~\ref{S:ApplyingFramework}; Section~\ref{S:ExpSetup} details our experimental setup, and Section~\ref{S:ResAndDisc} presents and discusses our results; finally, we provide conclusions and discuss future avenues of research in Section~\ref{S:ConclusionFuturework}.

\section{Radio-interferometric imaging}
\label{S:RadioInterferometry}
Radio-interferometers measure the sky using arrays of antennas (i.e. aperture arrays). Pairs of antennas (i.e. baselines) produce visibilities, the correlated instrumental response of the electrical field for some given time duration and electro-magnetic frequency. These can be related to the sky radiance distribution $I(l,m)$ using the measurement equation~\citep{smirnov2011revisiting}:
\begin{equation}
\label{Eq:RIME}
\begin{aligned}
& \text{V}(u, v) = C_{uv}\iint D_{uv}(l, m) \frac{I(l, m)}{n} e^{-2\pi i(ul + vm + w(n - 1))}\, dl\, dm, \\
& n = \sqrt{1 - l^2 - m^2}
\end{aligned}
\end{equation}
where $C$ represents the direction-independent effects, such as antenna gain, $D$ denotes the direction-dependent effects, such as phase gradients caused by the Earth's ionosphere, $(u, v, w)$ is the difference between antenna coordinates in the frame where $w$ is aligned with the phase center, and $(l, m)$ are the spatial angular coordinates on the celestial sphere, which is also the domain of the integral.

If we ignore the $D$ and $e^{-2\pi iw(n-1)}$ terms, Equation~\ref{Eq:RIME} simplifies to a two-dimensional Fourier transform, allowing us to retrieve an image of the sky emission through its inversion. This image contains artifacts due to the partial sampling caused by the nature of the antenna arrays and, especially in the case when imaging larger fields of view, by the omission of the $w$ and $D$ terms. Radio-interferometric imaging aims to correct for these, reconstructing an image that is usable for science.

\begin{figure}[t]
\centering
\begin{tikzpicture}[auto, node distance=2cm, >=latex]
    \node [name=visibilities] {};
    \node [draw, minimum height=3em, minimum width=6em, right of=visibilities] (deggrid) {De/gridding};
    \node [draw, minimum height=3em, minimum width=6em, right of=deggrid, node distance=4cm] (deconv) {Estimation};

    \draw [->] (deggrid) -- node[name=u] {$\tilde{\imath}^{\,n}$} (deconv);
    \coordinate [below of=u] (tmp);

    \draw [->] (visibilities) -- node {$\text{v}$} (deggrid);
    
    \node [draw, minimum height=3em, minimum width=6em, below of=u] (imup) {Image update};
    
    \draw [->] (deconv) |-  node[right] {$\bar{\imath}^{\,n}$} (imup);
    \draw [->] (imup) -| node[left] {$\hat{\imath}^{\,n+1}$} (deggrid);

\end{tikzpicture}
\caption{A high-level overview of the radio-interferometric pipeline. A reconstructed image $\hat{\imath}^{\,n}$ is compared to the measured visibilities $\text{v}$ in the de/gridding step, which outputs the residual $\tilde{\imath}^{\,n}$ in the spatial domain. This is passed to the estimation algorithm which generates $\bar{\imath}^{\,n}$, an estimate of $\tilde{\imath}^{\,n}$ without the effects of the measurement operator $F^\dagger G G^\dagger F$. The current reconstructed image $\hat{\imath}^{\,n}$ is then updated to $\hat{\imath}^{\,n+1}$ by adding $\bar{\imath}^{\,n}$ to it. As the estimation algorithm is usually iterative, the overall pipeline often has a major-minor loop structure.}
\label{Fig:RIPipelineOverview}
\end{figure}
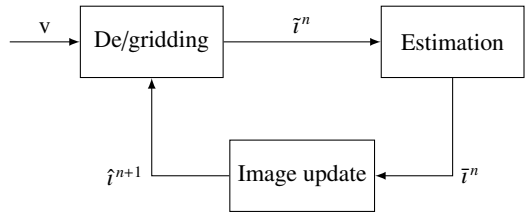

The general approach of most radio-interferometric imaging algorithms is iterative and is illustrated in Figure~\ref{Fig:RIPipelineOverview}. The residual $\tilde{\imath}^{\,n}$ between the measured visibilities $\text{v}$ and the current image estimate $\hat{\imath}^{\,n}$, also termed the dirty image when $n=0$, is computed in the de/gridding step. In image space, this can be expressed as
\begin{equation}
\tilde{\imath}^{\,n}= F^\dagger G(\text{v} - G^\dagger F\hat{\imath}^{\,n})
\label{Eq:MajorLoop}
\end{equation}
where $F$ and $F^\dagger$ are the Fast Fourier Transform (FFT) and its inverse respectively, and $G$ and $G^\dagger$ are the gridding and degridding operators for resampling visibilities to regular and irregular positions, respectively. There is a myriad of methods for $G$ and $G^\dagger$, most of which also include the correction for $w$, with some examples being the work of~\citet{cornwell2008noncoplanar, offringa2014wsclean, van2018image, ye2022high}.

The residual image $\tilde{\imath}^{\,n}$ is then sent to the estimation algorithm, which aims to remove the effect of the measurement operator $F^\dagger G G^\dagger F$ from the residual in a regularized manner, producing $\bar{\imath}^{\,n}$. Methods that aim to achieve this include CLEAN and its variants~\citep{hogbom1974aperture, cornwell2008multiscale, rau2011multi} that deconvolve $\tilde{\imath}^{\,n}$ in a greedy non-linear manner much akin to matching pursuit~\citep{Mallat1993}, and convex optimization methods based on Compressive Sensing~\citep{wiaux2009compressed, carrillo2014purify, dabbech2015moresane}. More recently, approaches that use Artificial Neural Networks have also been used, either in an end-to-end manner~\citep{Schmidt_2022, connor2022deep} or in plug-and-play~\citep{Terris_Dabbech_Tang_Wiaux_2022} or unfolded~\citep{Aghabiglou2024} frameworks. After estimation, the reconstructed image $\hat{\imath}^{\,n}$ is updated to $\hat{\imath}^{\,n+1}$ with
\begin{equation}
\hat{\imath}^{\,n+1}= \hat{\imath}^{\,n} + \bar{\imath}^{\,n}.
\label{Eq:ReconUpdate}
\end{equation}
The estimation step is usually iterative, therefore the imaging pipeline usually has a major-minor loop structure, with the loop shown in Figure~\ref{Fig:RIPipelineOverview} being the major, and the estimation being the minor.

De/gridding is often the bottleneck of the imaging pipeline, for example~\citet{tasse2018faceting} showed that it took up to 94\% of the total processing time for a serial implementation. Thus, there is much work that looks to expedite this step using parallelization. This can be done in a coarse or fine-grained manner. Fine-grained approaches parallelize on the local machine at a fine scale, such as per visibility or grid cell. There has been substantial amounts of work done in this area, both for the CPU as in~\citet{barnett2019parallel}, as well as the GPU as in~\citet{romein2012efficient, merry2016faster, veenboer2017}. 

Our method does not deal with fine-grained parallelization, and focuses primarily on coarser scales, aiming to parallelize the imaging pipeline across multiple nodes within a cluster. Typical methods for this are to parallelize according to the frequency and time axes. More recently, there has also been work that looks to distribute the gridding according to facets~\citep{monnier2022multi} as well as \chadded[id=rev]{by baseline~\citep{van2018image, pratley2019distributed} and} regions of the uv-plane~\citep{onose2016scalable, pratley2019distributed}. Different strategies can also be combined, such as in~\citet{gheller2023high} which parallelizes by time and the v-axis\chadded[id=rev]{, and the swiFTly framework~\citep{Wortmann_Kent_Nikolic_2024} which allows for simultaneous distribution both by facets in the spatial domain and subgrids in the Fourier domain}.

\chadded[id=rev]{Although similar in aim, our work differs from most of the aforementioned methods in that it takes into account the entire imaging process, including the deconvolution step, as opposed to only dealing with distributed de/gridding. To this end, the most similar methods to ours that are mentioned above } are those by~\citet{onose2016scalable} and~\citet{pratley2019distributed}\chdeleted[id=rev]{, }\chreplaced[id=rev]{. These works employ primal-dual methods for the image reconstruction, and distribute and parallelize by regions of the uv-grid, or also by baseline length in the case of~\citet{pratley2019distributed}}{both which distribute and parallelize by regions of the uv-grid, or also by baseline length in the case of~\citet{pratley2019distributed}}. Contrary to these approaches, which aim to solve directly for the measurement operator $F^\dagger G G^\dagger F$\chdeleted[id=rev]{, specifically in the case of primal-dual algorithms}, our work focuses on proposing a general framework that can be applied to a variety of different methods, especially in the traditional major-minor loop paradigm where convolution by the point spread function (PSF) is used as a surrogate for the measurement operator. 

\chadded[id=rev]{This provides several advantages over previous approaches. First, the general nature of our framework means that it is agnostic to the specific de/gridding and deconvolution algorithms used within. Thus, our proposed framework can be used in conjunction with many of the aforementioned methods. For example, for de/gridding, we currently use the improved w-gridder~\citep{ye2022high}, but we could instead use the image-domain gridder~\citep{van2018image} for a more fine-grained by-baseline parallelism per node, or use the swiFTly framework~\citep{Wortmann_Kent_Nikolic_2024} if we are applying our framework to a facet-based framework to take advantage of its distributed Fourier transform properties. Similarly, for deconvolution, we study our framework both with multi-scale CLEAN~\citep{cornwell2008multiscale} and a sparsity regularized convex optimization method akin to~\citet{wiaux2009compressed, carrillo2014purify, garsden2015lofar}, but can also look to employ more recent machine learning approaches~\citep{Terris_Dabbech_Tang_Wiaux_2022, Aghabiglou2024}. Second, presenting our framework in the traditional major-minor loop paradigm has performance benefits over approaches such as~\citet{onose2016scalable, pratley2019distributed} as de/gridding needs to only be performed every major cycle, rather than every iteration of the optimization algorithm.}

\chdeleted[id=rev]{This is beneficial as it implies that existing pipelines can be refactored to use our framework. It also has performance benefits as, rather than performing the costly de/gridding every iteration of the optimization algorithm, we only need to perform it every major cycle, of which there should be far fewer. }Finally, our framework is also decentralized, as opposed to the more centralized structure \chreplaced[id=rev]{of many of the previous works, where the gridded visibilities need to be collected in a central node for deconvolution}{of the work in~\citet{onose2016scalable, pratley2019distributed}, meaning that communication between nodes is not centralized, and no central node is required to perform the reduction}. Although we do not explicitly investigate this aspect of our framework in this paper, as we only study the reduced two-partition case, it may have performance implications when scaling to more partitions and larger datasets, as it allows for more local communication between nodes.

Other works that bear resemblance to ours are those of~\citet{Cai_Pratley_McEwen_2019} and~\citet{wang2024multi}. \citet{Cai_Pratley_McEwen_2019} reconstruct the image progressively by incorporating the missing information as previously reconstructed images. Their idea is similar to ours, but it differs in aim, as their goal is to derive an on-line reconstruction method to alleviate memory costs, whereas we look to improve scalability through parallelization. To this end, they partition in time, whereas we partition by spatial frequency. The work in this paper extends the framework presented in~\citet{wang2024multi} by introducing decentralized parallelism, as the previous framework only allowed for the partitioning of visibilities by spatial frequency, but was still serial. This paper also provides multi-scale CLEAN parallelization results, whereas the focus was previously only on sparsity regularized convex reconstruction.

\section{Decentralized framework}
\label{S:Parallelization}

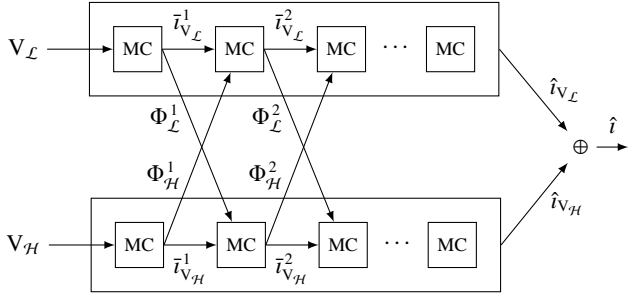
\begin{figure}[t]
\centering
\begin{tikzpicture}[auto, node distance=2cm, >=latex]
\node (low_vis) at (0, 0) [above=10pt]{$\text{V}_\mathcal{L}$};

\node (mc1l) at (low_vis.east) [draw, minimum height=0.75cm, minimum width=0.8cm, scale=0.8, right=25pt] {MC};
\node (mc2l) at (mc1l.east) [draw, minimum height=0.75cm, minimum width=0.8cm, scale=0.8, right=20pt] {MC};
\node (mc3l) at (mc2l.east) [draw, minimum height=0.75cm, minimum width=0.8cm, scale=0.8, right=20pt] {MC};
\node (low_ell) at (mc3l.east) [minimum height=0.75cm, minimum width=0.75cm, right=1pt] {$\cdots$};
\node (mcnl) at (low_ell.east) [draw, minimum height=0.75cm, minimum width=0.8cm, scale=0.8, right=0pt] {MC};

\draw [->] (low_vis) edge (mc1l);
\draw [->] (mc1l) -- node [midway, above=0pt] {$\bar{\imath}_{{\text{V}_\mathcal{L}}}^{\,1}$} (mc2l); 
\draw [->] (mc2l) -- node [midway, above=0pt] {$\bar{\imath}_{{\text{V}_\mathcal{L}}}^{\,2}$} (mc3l); 

\node[fit=(mc1l)(mc2l)(mcnl), draw, minimum height=0.75cm, minimum width=0.8cm, inner sep=9pt](vlnode){};

\node (full_vis) at (low_vis.south)[below=60pt]{$\text{V}_\mathcal{H}$};

\node (mc1h) at (full_vis.east) [draw, minimum height=0.75cm, minimum width=0.8cm, scale=0.8, right=25pt] {MC};
\node (mc2h) at (mc1h.east) [draw, minimum height=0.75cm, minimum width=0.8cm, scale=0.8, right=20pt] {MC};
\node (mc3h) at (mc2h.east) [draw, minimum height=0.75cm, minimum width=0.8cm, scale=0.8, right=20pt] {MC};
\node (low_ell) at (mc3h.east) [minimum height=0.75cm, minimum width=0.75cm, right=1pt] {$\cdots$};
\node (mcnh) at (low_ell.east) [draw, minimum height=0.75cm, minimum width=0.8cm, scale=0.8, right=0pt] {MC};

\draw [->] (full_vis) edge (mc1h);
\draw [->] (mc1h.east) -- node [midway, below=0pt] {$\bar{\imath}_{{\text{V}_\mathcal{H}}}^{\,1}$} (mc2h); 
\draw [->] (mc2h.east) -- node [midway, below=0pt] {$\bar{\imath}_{{\text{V}_\mathcal{H}}}^{\,2}$} (mc3h); 

\draw [->] (mc1h.east) -- node [midway, below left, xshift=-3pt, yshift=2pt] {$\Phi_\mathcal{H}^{\,1}$} (mc2l); 
\draw [->] (mc2h.east) -- node [midway, below left, xshift=-3pt, yshift=2pt] {$\Phi_\mathcal{H}^{\,2}$} (mc3l); 
\draw [->] (mc1l.east) -- node [midway, above left, xshift=-3pt, yshift=-2pt] {$\Phi_\mathcal{L}^{\,1}$} (mc2h); 
\draw [->] (mc2l.east) -- node [midway, above left, xshift=-3pt, yshift=-2pt] {$\Phi_\mathcal{L}^{\,2}$} (mc3h); 

\node[fit=(mc1h)(mc2h)(mcnh), draw, minimum height=0.75cm, minimum width=0.8cm, inner sep=9pt](vhnode){};

\node[fit=(mcnl)(mcnh)](centeringnode){};
\node (comb) at (centeringnode.east)[right=30pt]{$\oplus$};
\node (dummyout) at (comb.east)[right=12pt]{};

\draw [->] (vlnode.east) -- node [midway, right=3pt] {$\hat{\imath}_{\text{V}_\mathcal{L}}$} (comb); 
\draw [->] (vhnode.east) -- node [midway, right=3pt] {$\hat{\imath}_{\text{V}_\mathcal{H}}$} (comb); 
\draw [->] (comb) -- node [midway, above=2pt] {$\hat{\imath}$} (dummyout); 

\end{tikzpicture}
\caption{An illustration of our decentralized framework in the context of two partitions $\text{V}_{\mathcal{L}}$ and $\text{V}_{\mathcal{H}}$. We reconstruct two full-resolution images in parallel, achieved by sharing the \chadded[id=rev]{filtered} deconvolved residuals, \chadded[id=rev]{$\Phi_\mathcal{H}^n$ and $\Phi_\mathcal{L}^n$}, between different nodes after each major cycle, denoted as MC in the figure. \chadded[id=rev]{The deconvolved residuals are filtered both for normalization purposes, as well as to ensure that only spatial frequencies from their corresponding domains are present.} Low and high-resolution deconvolved images are produced after the first major cycle, denoted as $\bar{\imath}_{{\text{V}_\mathcal{L}}}^{\,1}$ and $\bar{\imath}_{{\text{V}_\mathcal{H}}}^{\,1}$ respectively. Afterwards, full-resolution deconvolutions $\bar{\imath}_{\text{V}_{\mathcal{L}}}^{\,n}$ and $\bar{\imath}_{{\text{V}_\mathcal{H}}}^{\,n}$, where $n>1$, are produced. The final reconstructed images of their respective nodes, $\hat{\imath}_{\text{V}_\mathcal{L}}$ and $\hat{\imath}_{\text{V}_\mathcal{H}}$, are then combined to produce the final reconstructed image $\hat{\imath}$.}
\label{Fig:InterleavedMultiStepPipeline}
\end{figure}

We propose a general framework for decentralized coarse-grained parallelization by spatial frequency of the radio-interferometric imaging pipeline. Each node reconstructs an image using only the local visibilities during the first major cycle. These images feature only partial spatial frequency information. They are then sent to all other nodes from the second major-cycle onward, allowing all nodes to reconstruct full-resolution images henceforth. The final reconstructed image can be obtained by computing the weighted-average of these full-resolution images. One way to view our framework is that we perform de/gridding in parallel, while creating pseudo full resolution problems to be solved by each node, using previously reconstructed images from other nodes as a surrogate for the missing visibilities. \chadded[id=min]{Note that this does result in a one major-cycle delay between the residual image of the local visibilities, and the received images from other nodes, which results in full-resolution images only being reconstructed after two major-cycles.}

For this paper, we focus on presenting and studying our framework within a reduced two-partition context, $\text{V}_\mathcal{L}$ and $\text{V}_\mathcal{H}$, which contain the low and high spatial frequency visibilities respectively, and \chreplaced[id=min]{are}{is} illustrated in Figure~\ref{Fig:InterleavedMultiStepPipeline}. \chadded[id=rev]{Here, after each major cycle $n$, the nodes corresponding to the partitions $\text{V}_\mathcal{L}$ and $\text{V}_\mathcal{H}$ send their latest filtered deconvolved residuals $\Phi_\mathcal{H}^n$ and $\Phi_\mathcal{L}^n$ (defined in Section~\ref{S:ApplyingFramework}), respectively, to the other node. These images are then used as a surrogate for the missing visibility information\chdeleted[id=amin]{, allowing each node to reconstruct full-resolution images from $n>1$ onwards}. The deconvolved residuals are filtered both for normalization purposes, as well as to ensure that only spatial frequencies from their corresponding domains are present, as the remaining spatial frequency information is available in the receiving node's locally stored visibilities.}

\begin{figure}[t]
\centering
\includegraphics[width=200pt]{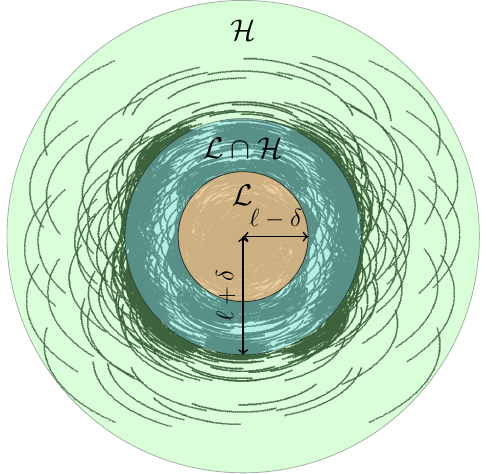}
\caption{The full set of visibilities is partitioned according to domains $\mathcal{L}$ (orange), which contains the low spatial frequency visibilities, and $\mathcal{H}$ (green), which contains the high spatial frequency visibilities. These are not mutually exclusive, but rather have an overlap region, which is denoted in cyan, and is defined using $\ell$, which is the radius of a circle bisecting $\mathcal{L} \cap \mathcal{H}$, and $\delta$, the half-width of $\mathcal{L} \cap \mathcal{H}$.}
\label{Fig:Partitioning}
\end{figure}

The partitioning of the visibilities is done according to their spatial frequency, which we determine using their uv-distance to zero, and each subset of visibilities is associated to a given compute node. Figure~\ref{Fig:Partitioning} illustrates an example partition configuration, where the low-resolution visibilities $\text{V}_\mathcal{L}$ belong to the orange domain $\mathcal{L}$, and the high-resolution visibilities $\text{V}_\mathcal{H}$ belong to the green domain $\mathcal{H}$. We set $\mathcal{L} \cap \mathcal{H} \neq \emptyset$ as de/gridding can cause frequency spillage after partitioning. We define the partitioning configurations using variables $\ell$ and $\delta$, with $\ell$ being the radius of a circle bisecting $\mathcal{L} \cap \mathcal{H}$, and $\delta$ being the half-width of $\mathcal{L} \cap \mathcal{H}$.

\section{Applying our decentralized framework to existing reconstruction methods}
\label{S:ApplyingFramework}
The proposed decentralized strategy is general and can be applied to different major-minor loop reconstruction methods. We apply our framework to two existing reconstruction algorithms. One is based on convex optimization with sparsity regularization~\citep{carrillo2014purify, garsden2015lofar}, and the other is multi-scale CLEAN~\citep{cornwell2008multiscale}. 

\subsection{Sparsity regularized convex reconstruction}
\label{SS:SparseConvRecon}
The sparsity regularized convex reconstruction method we apply within our framework produces a deconvolved residual $\bar{\imath}^{\,n}$ for every major cycle $n$. It achieves this by using FISTA \citep{Beck2009}  to solve:
\begin{equation}
\label{Eq:DeconvAllBaselines}
\begin{aligned}
\alpha^n &= \arg\min_{\alpha} \|\tilde{\imath}^{\,n} - H W\alpha\|_2^2 +  \lambda^n\|\alpha \|_1,\\
\bar{\imath}^{\,n} &= W \alpha^n
\end{aligned}
\end{equation}
where $\tilde{\imath}^{\,n}$ is the current residual image, $H$ the convolution operator by the PSF, $\lambda^n$ the current major cycle regularization parameter, and $W$ a wavelet transform operator. The final reconstructed image after $N$ major cycles is then $\sum_{i=1}^N \bar{\imath}^{\,i}$.

We previously studied this method in the context of a multi-step reconstruction that partitioned by spatial frequency~\citep{wang2024multi}. To apply our decentralized framework to this method, we augment the objective function in a similar way, with an additional data-fidelity term incorporating the received images from the other node, from the second major-cycle onward. To illustrate, the problem for the $\text{V}_\mathcal{L}$ node becomes:
\begin{equation}
\label{Eq:DeconvPartialBaselines}
\begin{aligned}
&\alpha_{\text{V}_{\mathcal{L}}}^{\,n} = \arg\min_{\alpha} \| \chreplaced[id=amin]{\Gamma_\mathcal{L}}{G_\mathcal{L}} (\tilde{\imath}_{\mathcal{L}}^{\,n} - H_\mathcal{L} W\alpha)\|_2^2 \\ & \qquad \qquad \qquad + \gamma_n\|\chdeleted[id=rev]{G_\mathcal{H}(}\chreplaced[id=min]{\rho^{n-1}_\mathcal{H}}{h^{n-1}} - \chadded[id=amin]{\Gamma_\mathcal{H}}W\alpha\chdeleted[id=rev]{)}\|_2^2 +\lambda_{\text{V}_{\mathcal{L}}}^{\,n}\|\alpha \|_1, \\
&\bar{\imath}_{\text{V}_{\mathcal{L}}}^{\,n} = W\alpha_{\text{V}_{\mathcal{L}}}^{\,n}, \\
&\chreplaced[id=min]{\rho^{n-1}_\mathcal{H}}{h^{n-1}} = \sum^{n-1}_{j=1} \chadded[id=rev]{\Phi_\mathcal{H}^j} - \chadded[id=amin]{\Gamma_\mathcal{H}} \sum^{n-1}_{j=1} \bar{\imath}_{\text{V}_{\mathcal{L}}}^{\,j},\\ 
&\chadded[id=amin]{\Phi_\mathcal{H}^j = \Gamma_\mathcal{H} \bar{\imath}_{\text{V}_{\mathcal{H}}}^{\,j},}\\
&\gamma_n  = 0\text{ if } n = 1 \text{ and } \gamma_n  = 1 \text{ otherwise }
\end{aligned}
\end{equation}
\chadded[id=amin]{where the subscripts $\mathcal{L}$ and $\mathcal{H}$ denote operators and images restricted to the domains of $\mathcal{L}$ and $\mathcal{H}$, respectively, and the subscript $\text{V}_\mathcal{L}$ refers to variables that are related to the node $\text{V}_\mathcal{L}$, but not necessarily restricted to its domain (e.g. $\tilde{\imath}_{\mathcal{L}}^{\,n}$ contains only spatial frequencies from $\mathcal{L}$, whereas $\bar{\imath}_{\text{V}_{\mathcal{L}}}^{\,n}$ is a full-resolution image produced by the $\text{V}_{\mathcal{L}}$ node).} 

\chreplaced[id=amin]{The}{where the} first data-fidelity term evaluates against the local visibility information $\text{V}_\mathcal{L}$ with PSF $H_\mathcal{L}$ and is akin to the data-fidelity term in Equation~\ref{Eq:DeconvAllBaselines}, and the second data-fidelity term contains the received reconstructed images condensed into the term $\chreplaced[id=min]{\rho^{n-1}_\mathcal{H}}{h^{n-1}}$, which subtracts the \chadded[id=rev]{high spatial frequencies of the} (already reconstructed) previous image estimate of the $\text{V}_\mathcal{L}$ node $\chadded[id=amin]{\Gamma_\mathcal{H}} \sum^{n-1}_{j=1} \bar{\imath}_{\text{V}_{\mathcal{L}}}^{\,j}$, from the \chadded[id=rev]{high spatial frequencies of the} previous image estimate of the $\text{V}_\mathcal{H}$ node, $\sum^{n-1}_{j=1} \chadded[id=rev]{\Phi_\mathcal{H}^j}$. This ensures that the previous image estimate of the $\text{V}_\mathcal{L}$ node gets subtracted from both fidelity terms. \chadded[id=amin]{We use the filters $\Gamma_\mathcal{L}$ and $\Gamma_\mathcal{H}$ ensure that the data-fidelity terms evaluate only their respective spatial frequencies, are normalized according to their respective variance, and that the weights in $\mathcal{L} \cap \mathcal{H}$ sum to one.} 

\chadded[id=min]{$\rho^{n-1}_\mathcal{H}$ can be viewed as a surrogate of Equation~\ref{Eq:MajorLoop} for the high-resolution spatial frequencies, evaluated without the measurement equation $FG^\dagger G F^\dagger$ as there are no locally stored visibilities at these frequencies. The first term is replaced by $\sum^{n-1}_{j=1}\Phi_\mathcal{H}^j$, as this is our current estimate of the sky at these spatial frequencies. The second term stays the same as Equation~\ref{Eq:MajorLoop}, where $\chadded[id=amin]{\Gamma_\mathcal{H}}\sum^{n-1}_{j=1} \bar{\imath}_{\text{V}_{\mathcal{L}}}^{\,j}$ is just the expanded version of the current filtered reconstructed image $\chadded[id=amin]{\Gamma_\mathcal{H}}\hat{\imath}_{\text{V}_\mathcal{L}}^{\,j}$.} 

\chdeleted[id=rev]{Finally, we use filters $G_\mathcal{L}$ and $G_\mathcal{H}$ to ensure that the data-fidelity terms evaluate only their respective spatial frequencies, are normalized according to their respective variance, and that the weights in $\mathcal{L} \cap \mathcal{H}$ sums to one.}

It should be noted that we do not explicitly compute $\chreplaced[id=min]{\rho^{n-1}_\mathcal{H}}{h^{n-1}}$, but rather update an aggregate image by adding the received and subtracting the locally deconvolved image every major cycle. This alleviates memory issues, particularly in the case of imaging large image cubes, as we are not required to keep copies of all the reconstructed images. \chadded[id=rev]{Additionally, we explicitly write out the terms for the computation of $\rho^{n-1}_\mathcal{H}$ within the sum in Equation~\ref{Eq:DeconvPartialBaselines} to emphasize that the received images $\Phi_\mathcal{H}^j$ are filtered by $\chadded[id=amin]{\Gamma_\mathcal{H}}$ before transfer.}

The $\text{V}_\mathcal{H}$ node solves the same problem, with the only difference being that the visibility partition and received image spatial frequencies are swapped for the first and second data-fidelity terms, i.e. we use $\tilde{\imath}_{\mathcal{H}}^{\,n}$ and $H_{\mathcal{H}}$ for the first data-fidelity term, and the residual image $\chreplaced[id=min]{\rho^{n-1}_\mathcal{L}}{l^{n-1}} = \sum^{n-1}_{j=1} \chadded[id=rev]{\Phi_\mathcal{L}^j} - \chadded[id=amin]{\Gamma_\mathcal{L}} \sum^{n-1}_{j=1} \bar{\imath}_{\text{V}_{\mathcal{H}}}^{\,j}$ for the second\chadded[id=rev]{, where $\Phi_\mathcal{L}^j = \chadded[id=amin]{\Gamma_\mathcal{L}} \bar{\imath}_{\text{V}_{\mathcal{L}}}^{\,j}$}.

\subsection{Multi-scale CLEAN}
The multi-scale CLEAN algorithm iteratively finds the brightest source at the appropriate scale from the residual image, and then updates the residual by subtracting the found source convolved by the PSF. For the $n_{th}$ major cycle, the deconvolved image produced can be denoted as:
\begin{equation}
\label{Eq:CLEAN}
\bar{\imath}^{\,n} = \text{ms-CLEAN}(\tilde{\imath}^{\,n}, H, \mathcal{S}, K) 
\end{equation}
where $\tilde{\imath}^{\,n}$ is the residual image, $H$ is the PSF, $\mathcal{S}$ the scales, and K the number of CLEAN iterations (i.e. minor cycles) .

Applying our decentralized framework to multi-scale CLEAN, the problem for the $\text{V}_\mathcal{L}$ node becomes:
\begin{equation}
\label{Eq:ParCLEAN}
\begin{aligned}
&\bar{\imath}_{\text{V}_\mathcal{L}}^{\,n} = \text{ms-CLEAN}(\tilde{\imath}_{(\mathcal{L} \cup \mathcal{H})}^{\,n}, H_{(\mathcal{L} \cup \mathcal{H})}^{\,n}, \psi^n, K), \\
&\tilde{\imath}_{(\mathcal{L} \chreplaced[id=min]{\Cup}{\cup} \mathcal{H})}^{\,n} = \gamma^n \chreplaced[id=amin]{\Gamma_\mathcal{L}}{G_\mathcal{L}} \tilde{\imath}_{\mathcal{L}}^{\,n} + \mu^n \chdeleted[id=rev]{G_\mathcal{H}} H_\mathcal{H} \chreplaced[id=min]{\rho^{n-1}_\mathcal{H}}{h^{n-1}}, \\
&H_{(\mathcal{L} \cup \mathcal{H})}^{\,n} = \gamma^n \chreplaced[id=amin]{\Gamma_\mathcal{L}}{G_\mathcal{L}} H_\mathcal{L} + \mu^n \chreplaced[id=amin]{\Gamma_\mathcal{H}}{G_\mathcal{H}} H_\mathcal{H}, \\
&\chreplaced[id=min]{\rho^{n-1}_\mathcal{H}}{h^{n-1}} = \sum^{n-1}_{j=1} \chadded[id=rev]{\Phi_\mathcal{H}^j} - \chadded[id=amin]{\Gamma_\mathcal{H}} \sum^{n-1}_{j=1} \bar{\imath}_{\text{V}_{\mathcal{L}}}^{\,j},\\ 
&\chadded[id=amin]{\Phi_\mathcal{H}^j = \Gamma_\mathcal{H} \bar{\imath}_{\text{V}_{\mathcal{H}}}^{\,j}.}\\
\end{aligned}
\end{equation}
Here multi-scale CLEAN deconvolves a dirty image $\tilde{\imath}_{(\mathcal{L} \chreplaced[id=min]{\Cup}{\cup} \mathcal{H})}^{\,n}$, which incorporates the high resolution information with $H_\mathcal{H} \chreplaced[id=min]{\rho^{n-1}_\mathcal{H}}{h^{n-1}}$, using the PSF $H_{(\mathcal{L} \cup \mathcal{H})}^{\,n}$. \chadded[id=min]{$\Cup$ denotes an asymmetric union operator, as $\tilde{\imath}_{(\mathcal{L} \Cup \mathcal{H})}^{\,n}$ contains all the spatial frequencies in $\mathcal{H}\cup\mathcal{L}$, but $\tilde{\imath}_{(\mathcal{L} \Cup \mathcal{H})}^{\,n} \neq \tilde{\imath}_{(\mathcal{H} \Cup \mathcal{L})}^{\,n}$ as the local visibilities are different for each node.} $\gamma^n$ and $\mu^n$ are equal to $1$ and $0$ respectively for $n=1$, and are otherwise weights corresponding to re-normalizing the PSF and dirty images\chadded[id=min]{ globally, as these are typically initially normalized based on their respective local visibilities}. $\psi^n$ is $\mathcal{S}_\mathcal{L} \subseteq \mathcal{S}$ for the $n=1$, which includes only scales present in $\mathcal{L}$, and is otherwise equal to $\mathcal{S}$. The rest of the operators and variables are the same as the ones described for Equation~\ref{Eq:DeconvPartialBaselines}. Likewise, the $\text{V}_\mathcal{H}$ node solves for a very similar problem, just with the various spatial frequency terms swapped, similar to Section~\ref{SS:SparseConvRecon}.

\section{Experimental setup}
\label{S:ExpSetup}
We evaluate our framework across two simulated and two real calibrated datasets. This section serves to provide a brief overview of these and how they are partitioned, as well as specify algorithmic parameters and details on the hardware and software.

\subsection{Datasets and partitioning}
\label{SS:Datasets}
\begin{figure}[t]
\centering
\includegraphics[width=220pt]{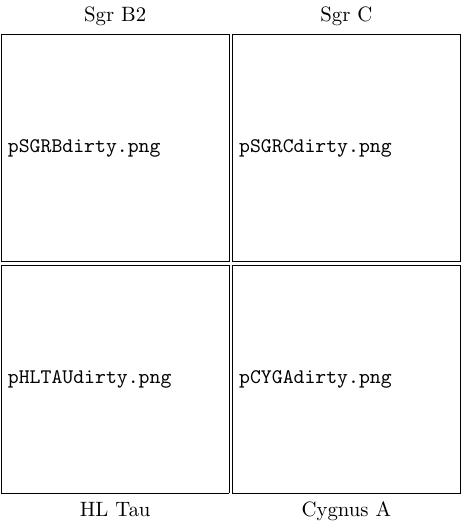}
\caption{Dirty images of the four datasets we use for our experiments. The top two are simulated, whereas the bottom two are real observations performed with ALMA and the VLA, respectively.}
\label{Fig:datasetdirties}
\end{figure}

\begin{table}[ht]
\centering
\caption{Dataset description}
\label{Table:Datasets}
\begin{tabular}{ c c c c c}
    \hline\hline
    Dataset & No. Vis & Resolution & Tel. \\
    \hline
    %Sgr A & 249600 & 512 $\times$ 512 @ 3.815'' & MeerKAT\\
    Sgr B2 & 9,361,440 & 512 $\times$ 512 @ 0.18'' & SKA-Mid\\
    Sgr C & 15,759,360 & 512 $\times$ 512 @ 0.429'' & SKA-Low\\
    HL Tau & 85,502,624 & 1500 $\times$ 1500 @ 0.005'' & ALMA\\
    Cygnus A & 82,246,592 & 1728 $\times$ 1728 @ 0.125'' & VLA\\
    \hline
\end{tabular}
\tablecomments{The resolution is first given in number of pixels in the image along each dimension, and then the angular resolution per-pixel in arc-seconds. The full SKA1 AA4 telescope configurations were used for generating the simulated data.}
\end{table}

Our simulated datasets consist of cutouts of regions around Sgr B2 and Sgr C from a Meerkat survey observed at 1.28GHz~\citep{heywood20221}. We simulate observations of these cutouts with pseudo RA-DEC coordinates, noise, and per-pixel angular resolutions, and use the telescope configurations of SKA1-Mid AA4 and SKA1-Low AA4, respectively. To create these datasets, we use RASCIL~\citep{cornwell_rascil} to generate the visibility positions with the given observation parameters, and obtain the visibility values by performing degridding on the ground-truth images with the improved w-stacking gridder~\citep{ye2022high}. 

Our real datasets are the HL Tau~\footnote[4]{\url{https://casaguides.nrao.edu/index.php?title=ALMA2014_LBC_SVDATA}} band 6 observation from the ALMA long-baselines survey and the first spectral window of the Cygnus A observation from~\cite{Sebokolodi_Cyga}. Together, our datasets account for a variety of different source types, uv-coverages, fields of view, and numbers of visibilities. The real datasets also account for possible calibration errors and realistic noise properties. We image all datasets using uniform weighting. Table~\ref{Table:Datasets} provides a high-level overview of our datasets, with their dirty images shown in Figure~\ref{Fig:datasetdirties}.

\begin{table}[t]
\centering
\caption{Number of visibilities per partition for test datasets}
\label{Table:PartitionConfigs}
\begin{tabular}{ c c c c c}
    \hline\hline
    Dataset & $\ell$ & $\text{V}_\mathcal{L}$ & $\text{V}_\mathcal{H}$ & $\text{V}_{\mathcal{L} \cap \mathcal{H}}$ \\
    \hline
    %Sgr A & 30 & 132648(.53) & 121140(.49) & 4188(.02)\\
    Sgr B2 & 20 & 6.99M(.75) & 2.53M(.27) & 0.16M(.02)\\
    Sgr C & 25 & 9.05M(.57) & 6.81M(.43) & 0.11M(.01)\\
    HL Tau & 60 & 39.86M(.47) & 46.24M(.54) & 0.61M(.01)\\
    Cygnus A & 40 & 41.65M(.51) & 41.83M(.51) & 1.22M(.01)\\
    \hline
\end{tabular}
\tablecomments{The number of visibilities are given in the millions, rounded off to the nearest 10k. The fraction of the partition sizes to the total dataset sizes are given in parentheses. The values for $\ell$ are in pixels.}
\end{table}

We separate each dataset into two partitions, looking to roughly balance the number of visibilities in each, as this should result in minimal waiting time between the nodes during reconstruction, with the number of visibilities per partition for each dataset shown in Table~\ref{Table:PartitionConfigs}. We found the Sgr B2 dataset to be particularly challenging to partition evenly. This is due to the high concentration of low-frequency visibilities present in the SKA1-Mid AA4 configuration. We discuss a possible solution to this issue in Section~\ref{SS:ReconSpeedAcc}. We set $\delta=1$ (pixel) when partitioning our datasets as we have previously shown in~\citet{wang2024multi} that the size of $\delta$ does not greatly change the final reconstruction quality, and it is best to keep $\mathcal{L} \cap \mathcal{H}$ small for efficiency.

\subsection{Algorithmic parameters}
For the sparsity regularized convex reconstruction, we use for $W$ a concatenation of the first 8 Daubechies wavelets and select $\lambda_n$, $\lambda_{\text{V}_{\mathcal{L}_n}}$ and $\lambda_{\text{V}_{\mathcal{H}_n}}$ using preliminary experiments. We found the general form of $\lambda_n = a\|\tilde{\imath}^{\,n}\|_2 \times \text{pow}(b,n)$ to work well for both the serial and parallel approaches, with $a$ and $b$ being dependent on the dataset. Generally, we found that it was necessary to increase $\lambda$ more aggressively in the $\text{V}_\mathcal{L}$ node when reconstructing using the parallel approach. We use 50 FISTA iterations as the stopping condition for both the serial and parallel implementations for all datasets. 

For multi-scale CLEAN, the scales are $\mathcal{S}=\{0, 1, 2, 4, 6, 10, 30\}$ for all our datasets, as we found that keeping the scales small is necessary to obtain reconstructions of similar resolution to the sparsity regularized convex reconstruction. We set the maximum number of CLEAN iterations to $K=2000$ per major-cycle, and set the absolute CLEAN threshold to values between 1e-4 and 2.5e-5 depending on the dataset. Unlike the typical approach, we do not convolve our model with the CLEAN PSF when evaluating the reconstructions, as this reduces the resolution of the final reconstruction, leading to poorer results. 

For purposes of comparison, we terminate all methods after five major-cycle iterations. This includes the case of multi-scale CLEAN, where the cleaning can terminate in fewer major iterations. The full-resolution reconstructions from each node are combined using equal weights: $\hat{\imath} = 0.5 \hat{\imath}_{\text{V}_\mathcal{L}}+0.5 \hat{\imath}_{\text{V}_\mathcal{H}}$ as preliminary experiments using inverse variance weighting produced very similar results. Finally, we don't add the residuals back into the reconstructed images for any of the evaluated methods.

\subsection{Implementation and hardware}
Our parallelization framework is implemented in python, using RASCIL~\citep{cornwell_rascil} for de/gridding, multi-scale CLEAN, and general quality of life functionality such as reading from measurement sets, and mpi4py~\citep{dalcin2005mpi} for the \chreplaced[id=min]{decentralized}{distributed} computing framework. We implemented the sparsity regularized convex reconstruction method in Julia~\citep{Julia-2017}, and integrated it into our python code. \chadded[id=rev]{We parallelized the wavelet transforms of this implementation as each individual wavelet in the over-redundant dictionary can be tranformed independently.}

We use a three-node architecture for our framework, using one \chreplaced[id=rev]{head}{master} and two \chreplaced[id=rev]{worker}{slave} nodes, with the \chreplaced[id=rev]{worker}{slave} nodes being responsible for the majority of the work, and the \chreplaced[id=rev]{head}{master} node responsible for aggregating statistics. In practice, this \chdeleted[id=min]{master-slave }architecture can be replaced with a decentralized one with minimal change, as our \chreplaced[id=rev]{head}{master} node is mainly idle and primarily serves to combine the individual reconstructed images and record the elapsed time. Our code and results are all available on our on-line repository, which will be provided upon publication.

We ran our experiments on the Jean Zay supercomputer~\footnote[5]{\url{http://www.idris.fr/eng/jean-zay/jean-zay-presentation-eng.html}} using the cpu\_p1 partition, which contains nodes each with 2 intel cascade lake 6248 CPUs and 192GB of primary memory. Our datasets are loaded from the \$WORK directory, which uses the performant parallel Lustre~\footnote[6]{\url{https://www.lustre.org/}} filesystem which manages a shared file system across multiple nodes. This should more closely reflect how disks are managed in the computer clusters of large radio interferometers, such as the SKA Science Data Processor.

\section{Results and discussion}
\label{S:ResAndDisc}
We evaluate the serial implementations of both multi-scale CLEAN and sparsity regularized convex reconstruction, and compare how they perform to their parallelized versions. For this, we primarily look to evaluate the reconstruction accuracy versus wall processing time. We also provide a more detailed breakdown and analysis of the division of computation time in order to provide a clearer picture of our framework's bottlenecks and possible areas to improve. \chadded[id=rev]{We also perform an experiment exploring how well our framework scales to larger images. For this, we provide a breakdown of computational times taken by the various subprocesses of our framework when imaging the Cygnus A dataset at the same field of view, but with a smaller angular pixel resolution, resulting in images of 10k $\times$ 10k pixels, and compare these to corresponding processing times obtained from imaging with the original field size of 1728 $\times$ 1728 pixels.}

\subsection{Reconstruction speed and accuracy}
\label{SS:ReconSpeedAcc}
We run all methods for five major-cycle iterations for each dataset, and plot the total processing time against reconstruction quality. As we have ground truths for our simulated datasets, we evaluate the quality of their reconstructions using signal-to-noise ratio (S/N) in decibels (dB), which we define as $\chreplaced[id=rev]{10}{20}\log\left(\dfrac{\|i\|_2}{\|i - \hat{\imath}\|_2}\right)$ where $i$ is the ground truth and $\hat{\imath}$ is the reconstructed image. 

For the real datasets, we instead compare the residuals of the final reconstructed images against a noise reference image that contains just the noise of the measurements. If the image is reconstructed perfectly, the reconstruction will contain only signal and the residual only noise, thus, the distribution between corresponding pixels of the final residual and noise reference should be the same. 

To obtain the noise reference image, we use a method based on jackknife resampling which has been previously applied to retrieve and study the noise and systematic properties of single-dish bolometric~\citep[\& references therein]{weiss2009large, romero2018multi} and interferometric~\citep[\& references therein]{di2023forming, vanMarrewijk2025} measurements in both the radio and mm domains. For testing the difference between the distributions, we specifically opt to use $\|W(r, n)_5\|_2$, where $r$ is the residual image and $n$ is the noise reference image. $W(x, y)_5$ here is the Wasserstein-1 distance~\citep{villani2009optimal} which produces an image of the same resolution of both $r$ and $n$, containing the minimum distance to move the one-dimensional distribution of the $5 \times 5$ window surrounding each pixel from $r$ to $n$. We opt for this over a statistical test that produces p-values, such as the Mann-Whitney U test~\citep{mann1947test}, as this differentiates between distributions that are far apart. We use the windowed approach as we only have one realization of the residual.

\begin{figure*}[t]
\centering
\includegraphics[width=500pt, clip]{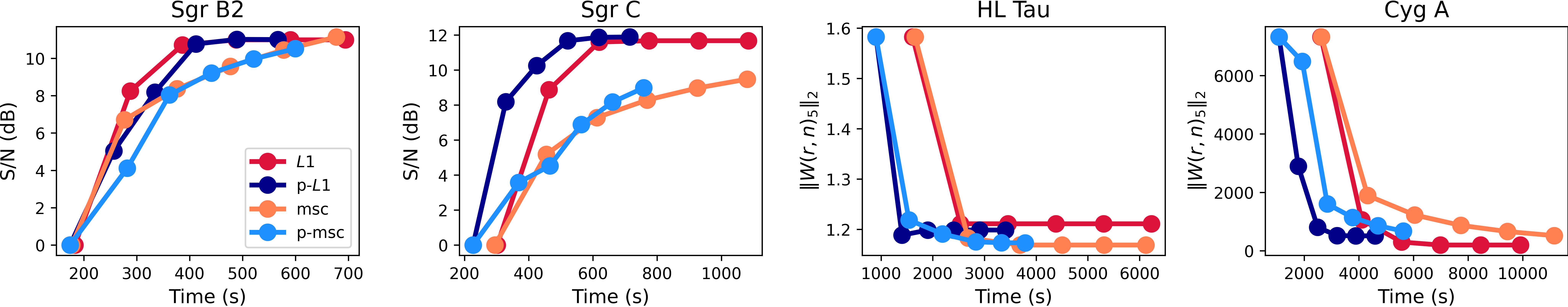}
\caption{\chadded[id=rev]{Time vs. accuracy for all four of our datasets, with each circle representing a major-cycle. The left two simulated datasets use S/N in dB for accuracy, whereas the two right real datasets use the $L^2$ norm of the per-pixel Wasserstein distances between the noisy reference image and the current residual, computed within a $5\times5$ sliding window as we only have one realization of the residual.}}
\label{Fig:TimeAcc}
\end{figure*}

Figure~\ref{Fig:TimeAcc} shows the results obtained for both our simulated and real datasets, with each major cycle demarcated with a circle, specifically just after the residual has been computed. The first data point of each plot corresponds to just after the initial dirty image was computed. In the interest of brevity, we refer to multi-scale CLEAN and the sparsity regularized convex reconstruction methods as msc and $L1$, respectively, with their parallel versions having a p- prepended. This is done both for our plots and our discussions of the methods henceforth.

We found that for smaller datasets, such as our simulated ones, the parallel methods do not show a large improvement compared to their serial counterparts, whereas they do for our larger datasets. There are two main reasons for this. The first is that the cost of deconvolution is substantial compared to the cost of de/gridding for the smaller datasets, resulting in less overall gain in speed. The second is due to the uneven partitions, such as in the case of our Sgr B2 dataset, which resulted in a lot of idling for the $\text{V}_\mathcal{H}$ node. As mentioned in Section~\ref{SS:Datasets}, we found it difficult to partition this dataset equally due to the extremely high density of visibilities in the center of the uv-plane. A possible remedy is to employ a method, such as baseline-dependent averaging~\citep{wijnholds2018baseline}, which will flatten the density distribution of visibilities in the uv-plane. 

Overall, we found that the p-$L1$ and $L1$ methods converge to images of similar quality. This is important as it shows that our framework does not introduce significant artifacts for these reconstruction methods, despite the problem being more challenging as the pseudo full resolution problems created for each node are less precise than if we were to reconstruct using all the visibilities in a single node. However, we did find p-$L1$ more difficult to regularize, with the $\text{V}_\mathcal{L}$ and $\text{V}_\mathcal{H}$ nodes sometimes requiring different regularization parameters. For multi-scale CLEAN, we found that for the same number of major-cycles, msc always reconstructed a slightly better image than the p-msc, which is largely expected due to the less precise nature of the pseudo full resolution problems. Despite this, we note that neither msc nor p-msc converged after our tested number of major-cycles \chadded[id=rev]{for most of our datasets, as we did not explicitly tune the number of minor cycles for each dataset}, thus, it is not clear whether the final converged images will have similar quality. We did find that in terms of equal-time reconstructions, p-msc generally reconstructs images that are of higher quality compared to msc, as illustrated in Figure~\ref{Fig:TimeAcc} for all our datasets except Sgr B2. \chadded[id=rev]{Finally, we do note that for the HLTau dataset where msc and p-msc do converge, the final image quality achieved by the two methods are similar. However, more experiments with a higher number of minor cycles are required to see if this result generalizes.}

A drawback to our decentralized framework is that it requires two major-cycle iterations to generate a full-resolution reconstruction. This means that, at least in our demonstrated case of two partitions, there is no large difference in reconstruction speed if the corresponding serial method can reconstruct most of the information in the first few major-cycles. This can be seen in some cases, such as in the Sgr C reconstruction, where the p-$L1$ and $L1$ methods converge in roughly the same amount of time, despite the faster major-cycles in p-$L1$. This will become less of an issue with larger datasets, more partitions, and more complex reconstructions where more major-cycle iterations are required.

\subsection{Division of computation time}
\label{SS:DivCompTime}
We divide the different sub-processes into six categories. \chreplaced[id=min]{The first three are degridding, gridding, and deconvolution}{The standard degridding, gridding, and deconvolution}, which refers to their in-memory operations (i.e. not including disk I/O)\chreplaced[id=min]{. Next we have disk read}{, as well as reading from disk}, which is performed every major cycle as we ingest datasets piecemeal one channel at a time during de/gridding to account for datasets that are too large to store in memory\chreplaced[id=min]{. Finally we have idle time and other. Idle time refers to the time that individual nodes spend waiting for the other nodes to synchronize, whereas other }{, and idle time. We also have a category for other, which }encompasses processes such as converting polarizations, memory allocations, computing weights, etc. These processing times are wall times obtained using clock queries within our code. 

Although our prototype implementation is largely unoptimized, we believe that this breakdown can nonetheless provide useful insight into the difference between the different methods and datasets as the same implementation is used for all our comparisons. Despite this, some aspects are bound to change with a more optimal implementation, for example we envisage the read portion to be drastically reduced.

\begin{figure*}[t]
\centering
\includegraphics[width=500pt]{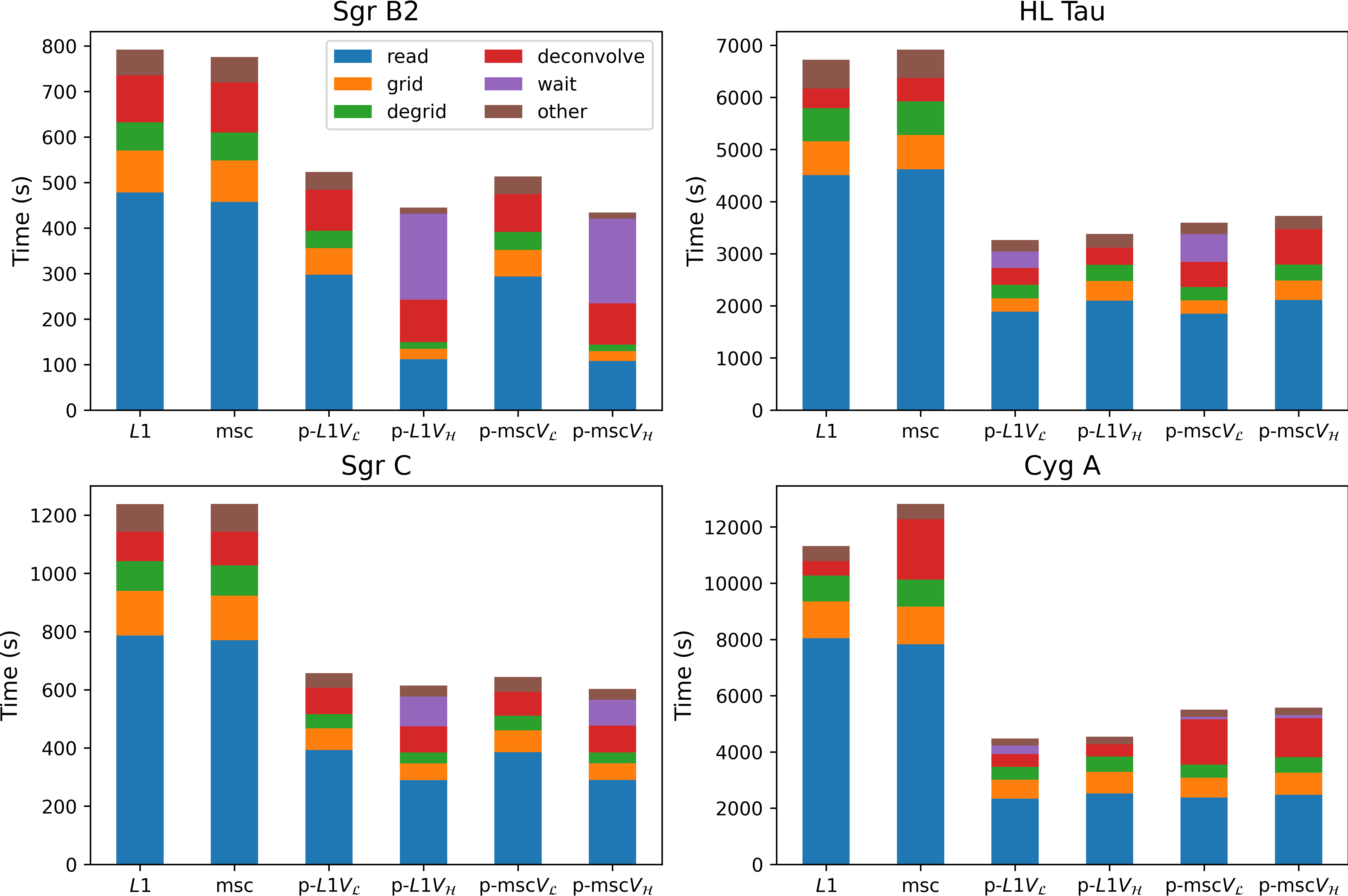}
\caption{\chadded[id=rev]{The breakdown of computation time taken by the various sub-processes for our four methods.}}
\label{Fig:TimeBreakdown}
\end{figure*}

Figure~\ref{Fig:TimeBreakdown} shows how much time each sub-process takes across the different methods for each dataset, with both $\text{V}_\mathcal{L}$ and $\text{V}_\mathcal{H}$ nodes being shown for the parallel methods. A result of note is that the reading takes up the lion's share of the de/gridding process, whereas the in-memory de/gridding often has comparable times to deconvolution. This is partially due to us reading all the visibilities from disk during each major-cycle, but we noticed after profiling with pidstat~\citep{sysstat} that most of this processing time was not spent directly on disk I/O but rather on overhead conforming the read visibilities to RASCIL's format. This overhead is also the primary culprit contributing to the anomalous result of the parallel methods demonstrating more than the optimal $2 \times$ speedup when applied to the Cygnus A dataset. A more efficient implementation should allow for computational speeds that are more comparable to current commonly used imaging software, such as WSClean~\citep{offringa2014wsclean}, and also allow us to obtain results for much larger datasets. \chadded[id=amin]{We also note that the overall conclusions of our results should still hold in this case, as both the de/gridding and the read subprocesses still increase with larger datasets.}

\begin{figure}[t]
\centering
\includegraphics[width=220pt]{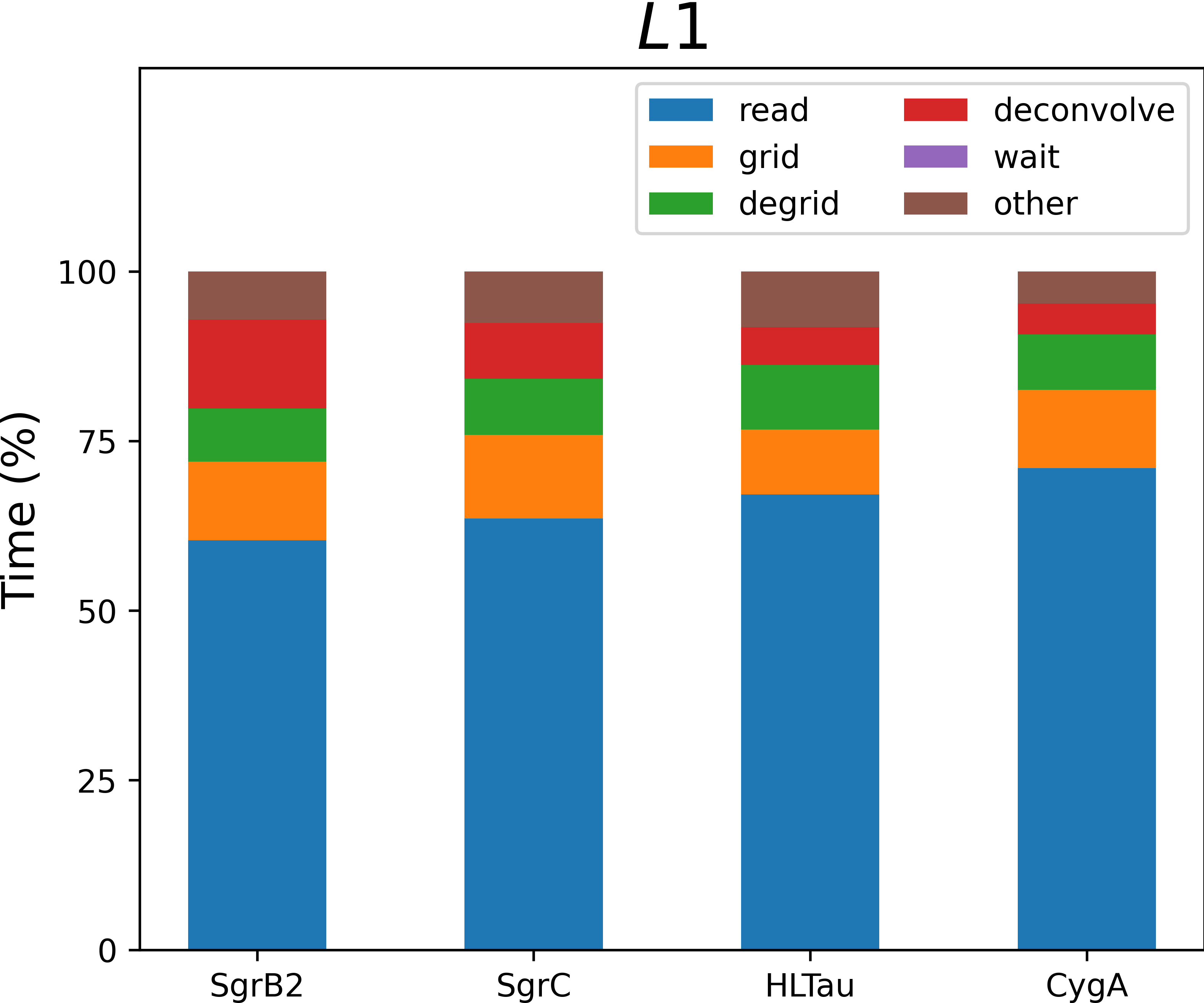}
\caption{\chadded[id=rev]{A percentage breakdown of processing time used by the various sub-processes of the $L1$ method across our four datasets.}}
\label{Fig:TimeBreakdownPerc}
\end{figure}

Finally, we note that although the deconvolution times are less pronounced with our larger datasets, they are by no means negligible, as shown in Figure~\ref{Fig:TimeBreakdownPerc} for the $L1$ method. This is especially the case with our parallel framework, where the cost of de/gridding is reduced per node. Thus, in order to achieve better scaling, it is important to not only look at increasing the number of partitions, but also methods for improving the computational cost of the deconvolution method. 

\subsection{Scaling to larger images}
\label{SS:ImageScaling}
\chadded[id=rev]{As our framework transfers images between nodes every major cycle, it is important to test how well it scales to imaging larger fields. For this, we image the Cygnus A dataset described in Section~\ref{S:ExpSetup} using p-msc and p-$L1$ at 10k $\times$ 10k pixels with a corresponding decrease in the pixel angular resolution, keeping the field of view the same. This results in the image sizes of 800MB, as opposed to roughly 24MB for the original test case of 1728 $\times$ 1728 pixels.}

\begin{table*}[ht]
\centering
\caption{\chadded[id=rev]{Cygnus A wall computation time breakdown}}
\label{Table:ImageScalingTimes}
\begin{tabular}{ c c c c c c c c c }
    \hline\hline
    Alg. & Node & Pix. res. & Deconv. & Degrid & Grid & Disk I/O & Transf. & Other \\
    \hline
    \multirow{4}{*}{p-msc} & $\text{V}_\mathcal{L}$ & 1728 $\times$ 1728 & 354.70s & 92.27s & 117.21s & 342.12s & 0.01s & 13.51s \\
    & $\text{V}_\mathcal{H}$ & 1728 $\times$ 1728 & 288.83s & 109.84s & 129.62s & 353.67s & 0.01s & 11.88s \\
    & $\text{V}_\mathcal{L}$ & 10k $\times$ 10k & 17778.44s ($\times$ 50) & 1450.58s ($\times$ 16) & 2519.88s ($\times$ 21) & 358.31s ($\times$ 1) & 0.45s ($\times$ 52) & 21.35s ($\times$ 2) \\
    & $\text{V}_\mathcal{H}$ & 10k $\times$ 10k & 18014.80s ($\times$ 62) & 2159.66s ($\times$ 20) & 2533.84s ($\times$ 20) & 362.50s ($\times$ 1) & 0.67s ($\times$ 51) & 21.11s ($\times$ 2) \\
    \hline
    \multirow{4}{*}{p-L1} & $\text{V}_\mathcal{L}$ & 1728 $\times$ 1728 & 91.66s & 92.44s & 111.52s & 332.31s & 0.02s & 11.52s \\
    & $\text{V}_\mathcal{H}$ & 1728 $\times$ 1728 & 89.33s & 108.17s & 126.85s & 359.43s & 0.02s & 13.49s \\
    & $\text{V}_\mathcal{L}$ & 10k $\times$ 10k & 3595.75s ($\times$ 39) & 1449.08s ($\times$ 16) & 2457.80s ($\times$ 22) & 365.53s ($\times$ 1) & 0.59s ($\times$ 30) & 20.16s ($\times$ 2) \\
    & $\text{V}_\mathcal{H}$ & 10k $\times$ 10k & 3573.73s ($\times$ 40) & 2173.30s ($\times$ 20) & 2555.44s ($\times$ 20) & 363.62s ($\times$ 1) & 0.60s ($\times$ 25) & 20.22s ($\times$ 1) \\
    \hline
\end{tabular}
\tablecomments{Wall computation times given are per major cycle, averaged across 4 major cycles (the first major cycle was omitted due to deconvolution times being very different from the rest due to only using a subset of the scales). The other category includes smaller tasks such as subtracting visibilities, adding images, memory allocation, applying weights, and converting polarization frames. The increase in time for the respective nodes between the different field sizes are provided in parentheses, rounded off to the nearest integer.}
\end{table*}

\chadded[id=rev]{Table~\ref{Table:ImageScalingTimes} shows a breakdown of the average wall processing times used by the various sub-processes for both the 10k $\times$ 10k and 1728 $\times$ 1728 configurations for a single major cycle, for both p-msc and p-L1. Overall, we found deconvolution to be the most expensive sub-process for both methods when increasing the pixel resolution. This was particularly the case with p-msc, where we found that RASCIL's fully serial implementation of the method scales poorly to larger image sizes, leading to its cost dominating all other subprocesses. Additionally, we also found a significant increase in de/gridding's processing time, although to a lesser extent when compared to deconvolution.}

\chadded[id=rev]{The increase in processing costs of both these sub-processes are roughly in line with the increase in number of pixels ($\approx33.5\times$) and are largely expected, as the deconvolution methods operate in image space, and the de/gridding algorithm has larger grids to convolve onto. This, in addition to the memory cost explosion, poses a problem if we wish to scale our framework to higher pixel resolutions, e.g. 100k  $\times$  100k, and necessitates that we investigate integrating our system with more scalable methods. For deconvolution, some examples are methods that take advantage of the sparsity in the domain of recovery, such as PolyCLEAN~\citep{jarretpclean}, and methods that separate the problem into sub-images, such as the DDFacet's SubSpace Deconvolution~\citep{tasse2018faceting}, Faceted HyperSARA~\citep{Thouvenin_2023}, or the parallel deconvolution of WSClean~\citep[\& references therein]{Wijnholds_2023}. As for de/gridding, we could look to reduce the grid size by splitting the de/gridding into a sum of smaller grids which we obtain by modulating the visibilities towards the center of the uv-plane, similar to the strategy used in image domain gridding~\citep{van2018image}.}

\chadded[id=rev]{We also found that the inter-node image transfer exhibited a significant increase in processing times, on a similar scale to deconvolution. However, it is currently not the bottleneck as its total processing time is negligible compared to the other subprocesses, taking slightly over half a second for the 10k $\times$ 10k case. As this operation is relatively straightforward, we can also predict its processing times when scaling to even larger images. For example, at the transfer rates shown in Table~\ref{Table:ImageScalingTimes}, which equates to roughly 1.3GB/s as the images are 800MB, a 100k $\times$ 100k image which has a size of 80GB will take around 65s to transfer, which is less than the cost of deconvolution of even the 1728 $\times$ 1728 case.}

\chadded[id=rev]{Although not an issue for our studied two-partition case, the transfer costs may become problematic once we extend our framework to more partitions. This is because a naive implementation would require each node sharing its deconvolved images with all other nodes every major cycle, resulting in transfer data amounts that grow quadratically with regard to the number of partitions. We envisage several methods that can mitigate this.}

\chadded[id=rev]{One is to employ a lossless compression scheme to the images before transfer. This should be achievable as the transferred images are band limited due to the supports of the filters $G_\mathcal{L}$ and $G_\mathcal{H}$ being finite in the Fourier domain. This also scales well to more partitions as the supports of the visibilities decrease with more partitions, and will be nearly mutually exclusive assuming a relatively small $\delta$ is used when partitioning the datasets. This would lead to near linear scaling, as opposed to quadratic. Another possible method that can mitigate transfer costs is to have more local communication, where nodes only communicate with a subset of the other nodes.}

\section{Conclusion and future work}
\label{S:ConclusionFuturework}
This paper proposes a decentralized radio-interferometric imaging framework that allows parallelizing by spatial frequency across two partitions. This additional axis of parallelization allows for more flexibility when scaling image reconstruction for future radio interferometers, such as the ones currently under construction for the SKA project. We applied our framework to two existing image reconstruction methods, and evaluated it against the serial implementations across four different data sets with various properties. We found that our framework enables faster reconstruction of images of similar quality compared to the serial implementations, and is particularly advantageous in the case of large data sets, such as the ones that will be produced by the SKAO telescopes, as the relative cost of deconvolution becomes less important.

The main drawback to our proposed method is that it requires at least two major cycles to converge. Thus, the performance advantages that it offers are less obvious if its serial counterpart converges in just a few major cycles. We believe this will not be an issue because the aim is to scale to very large, complex data sets that may require many major cycles to converge, and be separated into many partitions.

There are several avenues in which we are looking to expand our work: a) investigate our framework using more partitions, on larger datasets, which requires \chreplaced[id=rev]{both employing a compression method to the transferred images and defining a more intricate communication scheme in order to optimize communication costs}{defining the exchanges between nodes to optimize communication costs}, and a more intricate filter design, and is important to test both the scalability and the decentralized nature of our framework; b) apply our framework to state-of-the-art image reconstruction methods both to test how well it generalizes\chadded[id=rev]{, and to reconstruct larger fields in a scalable manner}; c) re-implement our prototype to be more performant, which will allow us to achieve more meaningful comparisons with other software and benchmark resource usage, such as CPU, memory, disk and network I/O, for larger datasets; d) investigate the use of baseline dependent averaging~\citep{wijnholds2018baseline} or a similar method with our framework, which should reduce the amount of idle time when using telescope configurations that have very high densities in the center of the uv-plane.

\section{Acknowledgements}
As part of the ``France 2030'' initiative, this work has benefited from a State grant managed by the French national research agency (Agence Nationale de la Recherche) attributed to the Exa-DoST project, and bearing the reference : ANR-22-EXNU-0004. This work was also financed by DARK-ERA (ANR-20-CE46-0001-01), UCA\textsuperscript{JEDI} (ANR-15-IDEX-01), and was granted access to the HPC resources of IDRIS under the allocation AD010415840 made by GENCI.

\bibliographystyle{aasjournal}
\bibliography{references}

\end{document}